# OBTAINING GLUON PROPAGATOR IN A NEW GENERALIZED GAUGE

**Jale Yılmazkaya Süngü**
Department of Physics, Kocaeli University - 41001 Izmit, Turkey
*jyilmazkaya@kocaeli.edu.tr*

**Arzu Türkan**
Özyeğin University, Department of Natural and Mathematical Sciences, Çekmeköy, Istanbul,Turkey
*arzu.turkan@ozyegin.edu.tr*

**Elşen Veli Veliev**
Department of Physics, Kocaeli University - 41001 Izmit, Turkey
*elsen@kocaeli.edu.tr*

**Abstract**

Gauge theories play a fundamental role in particle physics, nuclear physics and cosmology. The basic idea of these theories is that the Lagrangian density should be invariant under some transformations. Lagrangian invariance implies a certain freedom in defining gauge fields. In this study, the standard path integral quantization formalism is used. Gauge degrees of freedom manifest themselves in the difficulties in obtaining gauge field propagators. For a consistent quantization, it is necessary to eliminate non-physical gauge degrees of freedom. The standard procedure is to break the gauge symmetry by applying a gauge condition. In this work, we introduced a new generalized gauge condition $W_\mu A^{a\mu} = 0$, where $W_\mu = \lambda\, \partial_\mu + \beta\, n_\mu$, $n_\mu$ is an arbitrary constant four-vector, $\lambda$ and $\beta$ are real constant parameters. Using standard path integral quantization formalism, we obtained gluon propagators expressions in this new gauge. In the $\lambda = 1, \beta \to 0$, and $\beta = 1, \lambda \to 0$ limit cases, the obtained expression provides us the gluon propagators expressions available in the literature in covariant and non-covariant gauges, respectively. This study can give us a different perspective on quantization in Yang-Mills (YM) theories and can show the way to some ambiguities in quantum field theories (QFT).

# 1. INTRODUCTION

The idea of gauge fixing goes back to the Lorenz gauge approach to electromagnetism, which suppresses the excess degrees of freedom in the four-potential $A_\mu$ while preserving manifest Lorentz invariance. The gauge invariance of Maxwell's equations in the quantum form of the theory is directly related to invariance under a local phase transformation of the quantum fields. The origin of gauge invariance in classical electromagnetism emerges from the fact that the potentials $\vec{A}$ and $V$ are not unique for a given electric and magnetic fields. This requires eliminating of non-physical degrees of freedom taken from each other by gauge transformations corresponding to the same physical fields. The principle of gauge invariance is essential to constructing a workable quantum field theory. To perform a perturbative calculation in a gauge theory, we have to first fix the gauge, adding terms to the Lagrangian density resulting in the breaking of the gauge symmetry to eliminate these unphysical degrees of freedom.

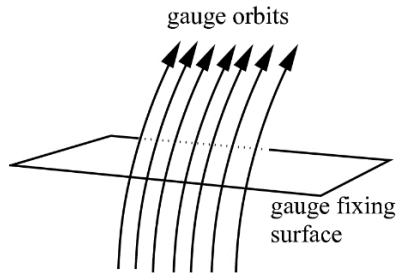

**Figure 1**. The constraint surface equation in functions space. The gauge fixing conditions must be intersects each gauge orbit exactly once. (Cline, 2020).

The concept of gauge invariance is also generalized for the theories of weak and strong interactions. For this reason, they are all known as "gauge theories". QED is a special case of gauge theory (Feynman, 1950), as it is related to an Abelian structure. The generalization to so-called non-Abelian gauge theories, or Yang-Mills theories (Yang-Mills, 1954), is necessary for particle physics, as they are a building block of the standard model. Even classical general relativity can be considered a gauge theory of related type, and thus will any quantum version of it. One should note that the physical observables do not rely on the gauge condition so every gauge choice should yield to the same physical results.

However, physicists had trouble calculating non-abelian amplitudes beyond the lowest order of perturbation theory. Then Faddeev-Popov solved this problem (Faddeev and Popov, 1967) for all orders using path-integral tricks. In other words, the standard way to quantize a gauge field theory is the Faddeev-Popov approach which is effectively achieved only if a given gauge condition is imposed on Lagrangian.

The widely used gauge conditions in the strong interactions are covariant gauges, such as the Landau and the Feynman and non-covariant ones, such as the Coulomb gauge. Non-covariant gauges (light-cone, axial, coulomb, planar, temporal, etc.) have been extensively employed in standard model calculations and string theories (Leibbrandt, 1994, Das and Frenkel, 2005, Suzuki and Sales, 2004). Axial and light-cone gauges are generally used in Quantum Chromo Dynamics (QCD) since its freedom from ghosts and is also helpful in treating the Chern-Simon theories. Nonetheless, the propagator in axial type gauge contains singularities. The treatment of these poles requires introducing non-covariant gauges (Veliev et al., 1989, Veliev, 2001). Light-cone gauge is preferable for the cancellation of anomalies in superstring theories. In Coulomb gauges, it appears as the problem of ill-defined energy integrals as the propagator for the time-like component. The coulomb gauge is utilized in the discussion of confinement, as well.

The paper is organized as follows. After a brief introduction to the quantization techniques in QFT in section 2, we extract the gluon propagator from quantization. In section 3, we present our results while the discussion is given in section 4.

## 2. DERIVATION OF GLUON PROPAGATOR FROM QUANTIZATION

There are two standard methods to transform the classical action into a QFT.

- One is *canonical formalism*, where the fields are treated as operators on a Fock space and canonical (anti-) commutation relations are imposed.
- The other is the *path-integral formalism*, where an integral over all fields is performed, which provides an intuitive picture of how quantum corrections augment the classical field configurations.

While gauge theories can be formulated in canonical quantization, this is at least cumbersome. Consequently, it is used little in practice, and not every issue has even been treated in this approach. The path integral is a much more natural framework for gauge theories. The path integral formulation is as axiomatic as canonical quantization, and it cannot be derived. Here we use path-integral formalism for quantization.

QCD is a theory that explains the interaction of quarks and gluons via color charge. It is defined by the Yang-Mills locale symmetry group in the non-abelian group $SU(3)$ in QCD. The free quark QCD Lagrangian is defined as;

$$L_{QCD} = \overline{\Psi}_j (i\gamma_\mu \partial_\mu - m_{jk}) \Psi_k, \tag{1}$$

here, $\gamma_\mu$ represent Dirac gamma matrix, $j, k$, color indices, and $\Psi_k, \overline{\Psi}_j$ quark and antiquark fields, respectively, $\mu, \nu = \overline{1,4}$ are Lorentz vector index, $m_{jk}$ is quark mass.

The Lagrangian $L_{QCD}$ is invariant under global ($x$–independent) gauge transformations

$$\psi(x) \rightarrow U\psi(x), \tag{2}$$

with unitary and unimodular matrices $U^+ = U^{-1}$, $|U| = 1$, belonging to the fundamental representation of the colour group *SU(3)*. The matrices $U$ can be represented as

$$U \equiv U(\theta) = \exp(i\theta^\alpha t^\alpha), \tag{3}$$

where $\theta^\alpha$ are the gauge transformation parameters. Invariance under the global gauge transformations in Eq. (2) can be extended to local ($x$-dependent) ones, i.e., to those where $\theta^\alpha$ in the transformation matrix in Eq. (3) is $x$-dependent. This can be achieved by introducing gluon fields $A_\mu^a(x)$, which transform according to

$$A_\mu(x) \rightarrow A_\mu^{(\theta)}(x) = U A_\mu(x) U^{-1} + i g (\partial_\mu U) U^{-1}, \tag{4}$$

and the partial derivative $\partial_\mu$ in Eq. (1) is replaced by the covariant derivative $D_{jk}^\mu$,

$$D^\mu_{jk} = \delta_{jk}\partial^\mu + ig_s(t_\alpha)_{jk}A^\mu_\alpha \tag{5}$$

with $g_s$ the strong coupling related to $\alpha_s$ by $g_s^2 = 4\pi\alpha_s$, $t_\alpha$ is the generator of Lie Algebra of the gauge group $SU(3)$. By convention, the constant of proportionality is normally taken to $t_\alpha = (1/2)\lambda_\alpha$ with $\lambda_\alpha$ being the Gell-Mann matrices, and $A^\mu_\alpha$ gluon fields, α colour index (i.e., α ∈ [1, . . . , 8]), antisymmetric strength tensor of gluon fields is also expressed as:

$$F^{\mu\nu}_\alpha = \partial^\mu A^\nu_\alpha - \partial^\nu A^\mu_\alpha - g_s f_{abc}A^\mu_b A^\nu_c. \tag{6}$$

here, $f_{abc}$ is the group structure constant in QCD. Since QCD has a non-abelian group under local symmetry, it has different features, such as "Asymptotic Freedom and confinement" that are not found in other theories (Muta, 1998).

The second step for studying QCD is the ground partition function of the theory reads:

$$Z[J^a_\mu] \equiv e^{iW[J^a_\mu]} = \int [D\phi]e^{S_{eff}}, \tag{7}$$

here, $c$ and $\bar{c}$ show the ghost fields, $[D\phi] = [DA][D\psi][D\bar{\psi}][Dc][D\bar{c}]$ denote the functional measure including all fields, and the effective action is:

$$S_{eff} = \frac{1}{2}\int d^4x \left[A^{\mu a}(x)\left(-\Box g_{\mu\nu} + \partial_\mu \partial_\nu\right)A^{\nu a}(x) + \frac{1}{\alpha}\left(W_\mu A^{\mu a}\right)f(W_\nu A^{\nu a})\right] \tag{8}$$

where $g_{\mu\nu}$ represents the metric tensor. The last term comes from the gauge fixing process.

The gluon propagator is defined as:

$$D^{ab}_{\mu\nu}(x-y) = \frac{\delta^2 \ln Z}{\delta J^a_\mu(x)\delta J^b_\nu(y)}, \tag{9}$$

here $J^a_\mu(x)$ is an external c-number source for the field $A^a_\mu(x)$. In Eq. (9) $D^{ab}_{\mu\nu}(x-y)$ can be expressed in terms of gluon self-energy tensor $\Pi_{\mu\nu}(k)$ by the Schwinger-Dyson equation:

$$D^{-1}_{\mu\nu}(k) = D^{-1}_{0\mu\nu}(k) + \Pi_{\mu\nu}(k), \tag{10}$$

here $D^{-1}_{0\mu\nu}(k)$ is the bare gluon propagator and $D^{-1}_{\mu\nu}(k)D_{\nu\lambda} = \delta_{\mu\lambda}$. The vacuum self-energy satisfies the transversality constraint $k^\mu k^\nu \Pi_{\mu\nu}(k) = 0$, which leads to the expression $\Pi_{\mu\nu}(k) = (g_{\mu\nu}k^2 - k_\mu k_\nu)\Pi(k^2)$, here $\Pi(k^2)$ is invariant function.

In this work, we define a unified gauge condition as;

$$W_\mu A^{\mu a} = 0 \quad \text{with} \quad W_\mu = \beta n_\mu + \lambda \partial_\mu. \tag{11}$$

Here, $A^{\mu a}$ Yang-Mills fields, $\beta$ and $\lambda$ are the coefficients. $n_\mu$ is an arbitrary constant 4-vector. Generalized gauge condition can be written in coordinate space as follows:

$$(\beta n_\mu + \lambda \partial_\mu) A^{\mu a} = 0. \tag{12}$$

In momentum space, the below expression can be employed as:

$$(\beta n_\mu + i\lambda k_\mu) A^{\mu a} = 0. \tag{13}$$

The inverse form of gluon propagator expression is:

$$D_{0\mu\nu}^{-1}(k) = k^2 g_{\mu\nu} - k_\mu k_\nu + \frac{1}{\alpha}(\beta n_\mu - i\lambda k_\mu) f (\beta n_\nu + i\lambda k_\nu). \tag{14}$$

We can write down Eq. (14) as;

$$D_{0\mu\nu}^{-1}(k) = k^2 g_{\mu\nu} - A k_\mu k_\nu + B n_\mu n_\nu + C n_\nu k_\mu + E n_\mu k_\nu \tag{15}$$

$$A = 1 - \frac{f}{\alpha}\lambda^2, \quad B = \frac{f}{\alpha}\beta^2,$$

$$C = -\frac{i\lambda f \beta}{\alpha}, \quad E = \frac{i\lambda f \beta}{\alpha}.$$

Here substituting $f = 1$ and using the fact that $D_{\mu\nu}^{-1}(k) D_{\nu\lambda} = \delta_{\mu\lambda}$, we have the gluon propagator:

$$D_{0\mu\nu}(k) = F g_{\mu\nu} + G k_\mu n_\nu + H k_\nu n_\mu + L k_\mu k_\nu + M n_\mu n_\nu, \tag{16}$$

where

$$F = \frac{1}{k^2}, \quad G = -\frac{B}{k^2[(n \cdot k)B + E k^2]}, \quad H = -\frac{E}{k^2[(n \cdot k)E + (1-A)k^2]},$$

$$L = \frac{1}{k^2} \frac{E C n^2 + A(1-A)k^2}{[(1-A)k^2 + C(n \cdot k)][(1-A)k^2 + E(n \cdot k)]}, \quad M = 0. \tag{17}$$

Finally, the following relation is obtained for the gluon propagator:

$$D_{0\mu\nu}^{ab}(k) = \frac{\delta^{ab}}{k^2}\left[ g_{\mu\nu} - \beta^2 \frac{k_\mu n_\nu + k_\nu n_\mu}{\beta^2 (n \cdot k)^2 + \lambda^2 k^4}(n \cdot k) + i\lambda\beta k^2 \frac{k_\mu n_\nu - k_\nu n_\mu}{\beta^2 (n \cdot k)^2 + \lambda^2 k^4} \right.$$
$$\left. + \frac{k_\mu k_\nu}{\beta^2 (n \cdot k)^2 + \lambda^2 k^4}(\beta^2 n^2 - \lambda^2 k^2 + \alpha k^2) \right], \tag{18}$$

where $\alpha$ is an arbitrary coefficient. In the limit $\beta \to 0$, if we substitute the $\lambda = 1$ in Eq. (18), so we get the below expression in the form of covariant gauges:

$$D_{0\mu\nu}^{ab}(k) = \frac{\delta^{ab}}{k^2}\left[g_{\mu\nu} - \frac{k_\mu k_\nu}{k^2}(1-\alpha)\right]. \tag{19}$$

In the $\beta = 1$, $\lambda \to 0$ condition, we can find out the gluon propagator in non-covariant gauges:

$$D_{0\mu\nu}^{ab}(k) = \frac{\delta^{ab}}{k^2}\left[g_{\mu\nu} - \frac{k_\mu n_\nu + k_\nu n_\mu}{(n \cdot k)} + \frac{k_\mu k_\nu (n^2 + \alpha k^2)}{(n \cdot k)^2}\right]. \tag{20}$$

In the end, explicit analytical expression is derived for the gluon propagator in a generic gauge condition.

## 3. RESULTS

In the context of the QCD theory, we derive the gluon propagator both covariant and non-covariant gauges from quantization proposing a general condition, $W_\mu A^{a\mu} = 0$ with $W_\mu = \lambda \, \partial_\mu + \beta \, n_\mu$. Employing standard path integral quantization method, gluon propagator expressions are evaluated in a generalized gauge. In the $\lambda = 1$, $\beta \to 0$, and $\beta = 1$, $\lambda \to 0$ limit cases, the obtained expressions are consistent with the results available in the literature in covariant and non-covariant gauges, respectively.

## 4. DISCUSSION

Understanding the quantization technique in QCD is crucial in order to solve some problems, such as Gribov ambiguities (Gribov, 1978) which discuss the Faddeev-Popov procedure is consistent only at the perturbative level. In this context, we have proposed a new gauge-fixing condition to study the behaviour of different Green functions in the framework of perturbative approach.

## 5. CONCLUSION

These studies like these allow us to understand the quantization process in a deep way. This study can give us a different perspective on quantization in Yang-Mills theories and can show the way to some ambiguities in quantum field theories.